%% file: nutmass.tex
\documentclass[12pt,twoside]{article}
\usepackage{def,physics,epsfig,cite}
 
\newlength{\capwidth}
\relax
\input{nutmass_include.tex}
\newcommand {\fignopsfac}{1}      
\newcommand {\figprobs}{2}        
\newcommand {\tabnolimits}{1}     
\begin{document}
\bibliographystyle{unsrt}
\begin{titlepage}
%
\noindent{\large\bf%
Northeastern University    \hspace*{\fill}   NUB / 3146  \\
Boston, MA 02115, USA      \hspace*{\fill}   \today} \\

\vbox{
\vspace{2.5em}
\bigskip
\bigskip
\bigskip}

\begin{center}
{\Large\bf    Constraints on the tau neutrino mass and mixing \vspace*{3mm}\\
              from precise measurements of tau decay rates}
\end{center} 
\vspace{2.0em}
{\normalsize
\begin{center}
{\large \bf J. Swain and L. Taylor}
\end{center} }
\vspace{1.5em}
{\normalsize
\begin{center}
{\large \bf Northeastern University} \\
{\large Boston, MA 02115, USA}\\
\end{center} }
\vspace{2em}
 
\vspace{3em}
\begin{center}
{\large\bf Abstract}
\vspace{0.5em}

\parbox{140mm}
{
\noindent We have derived constraints on the tau neutrino mass and fourth generation 
mixing from an analysis of the partial widths of tau lepton decays, in particular:
$\tau^-\rathin\mathrm{e}^-\bar{\nu}_{\mathrm{e}}\nutau$,
$\tau^-\rathin\mu^-\bar{\nu}_\mu\nutau$,
$\tau\rathin\pi^-\nutau$, and
$\tau\rathin\mathrm{K}^-\nutau$. 
We present predictions for the tau decay widths,
allowing for a non-zero tau neutrino mass, $m_{\nu_\tau}$, 
and for mixing with a neutrino of mass  
$m_{\nu_{\mathrm{L}}}>M_{\mathrm{Z}}/2$, which is parametrised 
using a Cabibbo-like mixing angle, $\theta_{\mathrm{L}}$.
By comparison of these theoretical predictions with the experimental measurements, 
we obtain the following bounds at the 90\% confidence level:  
\mbox{$m_{\nu_\tau}<\MNUTCvalintA$\,{\em{MeV}}} and 
\mbox{$\sin^2\theta_{\mathrm{L}}<\FMIXCvalintA$}. 
}
\end{center}
 
\end{titlepage}
%
\pagestyle{empty}
\mbox{ }
\clearpage
\pagestyle{plain}
\addtocounter{page}{-1}
\pagenumbering{arabic}
\section{Introduction}
Massive neutrinos feature in many extensions to the Standard Model\cite{MASSIVENU}.
They have been suggested as a source of dark matter\cite{DARKMATTER} and,
if they oscillate in the sun, 
as an explanation for the deficit in the observed solar neutrino 
flux\cite{SOLARDEFICITA,SOLARDEFICITB}.
The best experimental upper limit on the tau neutrino mass 
is $m_{\nu_\tau} < 24$\,MeV at the 95\% confidence level\cite{ALEPHNUTAU}, 
which was obtained using many-body hadronic decays of the $\tau$.
 
In this paper we describe a complementary method for constraining the $\tau$ neutrino
mass, $m_{\nu_\tau}$, from precise measurements of $\tau$ partial widths 
for the following decays%
\footnote{Henceforth we denote the branching ratios for these processes as
          $\Be, \Bm, \Bp, \Bk$ respectively;
          $\Bl$ denotes either $\Be$ or $\Bm$ while $\Bh$ denotes either $\Bp$ or $\Bk$.}
:%
$\tau^-\rathin\mathrm{e}^-\bar{\nu}_{\mathrm{e}}\nutau$,
$\tau^-\rathin\mu^-\bar{\nu}_\mu\nutau$,
$\tau^-\rathin\pi^-\nutau$, and
$\tau^-\rathin\mathrm{K}^-\nutau$. 
The dependence of the purely leptonic decay rates on $m_{\nu_\tau}$ has been considered by 
others\cite{SHROCK81B,BRYMAN87A,SAMUEL88A} 
whereas, to the best of our knowledge, the hadronic 
decays have not previously been analysed for this purpose.
The $\tau$ partial decay widths are also sensitive to mixing in the leptonic sector
with a fourth generation weak isospin lepton doublet $(\nu_{\mathrm{L}},{\mathrm{L}}^-)$
\cite{SHROCK81A,SHROCK81B,SHARMA84A,RAJPOOT88A,LI91A,MARCIANO95A},
where the neutrino has a mass $M>M_{\mathrm{Z}}/2$. 

We present calculations of the $\tau$ partial widths for these channels
which allow for the effects of non-zero tau neutrino mass and mixing.
Then, using data from $\mathrm{e^+e^-}$ experiments, 
we derive constraints on the tau neutrino mass and mixing. 

This approach is complementary to traditional analyses of the kinematic 
end-point of multi-hadron tau decays, which yield tighter constraints
on $m_{\nu_\tau}$ but which are insensitive to fourth generation mixing.
Moreover, the channels we consider are statistically independent and 
theoretically very well understood.
\section{Partial widths for $\mathbf{\tau}$ decays}
In the Standard Model describing the electroweak interaction,
the partial width $\Gammalept$ for the decay
$\tau^-\rathin\ell^-\bar{\nu}_{\ell}\nutau (X_{\mathrm{EM}})$, with
$\ell^-=\mathrm{e}^-, \mu^-$ and $X_{\mathrm{EM}} = \gamma,~\gamma\gamma,~\epem,\ldots$, is given by
\begin{equation}
  \Gammalept   \equiv   \frac {\Bl} {\tau_\tau}
                  =       \frac {\GF^2 m_\tau^5}   {192\pi^3} \radcorlept \psfaclept,
\label{equ:gammalept}
\end{equation}
where $\GF$ is the Fermi constant, $m_\tau$ and $\tau_\tau$ are the $\tau$ mass and lifetime.
The radiative--correction function $\radcorlept$ has
been calculated\cite{BERMAN58A,KINOSHITA59A,SIRLIN78A,MARCIANO88A},                          
with $\alpha(m_\tau)\simeq 1/133.3$\cite{MARCIANO88A}, to be
\begin{equation}
   \radcorlept = \left[ 1 - \frac{\alpha}{2\pi} \left( \pi^2 - \frac{25}{4} \right) \right]
                 \left[ 1 + \frac{3}{5} \frac{m_\tau^2}{m_W^2} + \cdots \right] 
                \simeq  0.9960.
   \label{equ:radcorlept}
\end{equation}
%
%
%
If the neutrino masses are zero for all generations,
then the phase-space factor, $\psfaclept^0$, is given by the well-known expression:
\begin{equation}
    \psfaclept^0 (x) 
        =     1 -  8x - 12 x^2{\mathrm{ln}}x + 8 x^3 - x^4 
      \simeq  \cases {1,      & ($\ell=\mathrm{e}$); \cr
                        0.9726, & ($\ell=\mu$);          \cr}
    \label{equ:psfaclept}
\end{equation}
where $x=m_\ell^2/m_\tau^2$.
%
A value of $\GF=\GFMUval$\cite{PDG96}                                                     
is obtained from the measurement of the muon lifetime
using equation \ref{equ:gammalept}, with $\ell^-=\mathrm{e}^-$ and substituting $\tau\rathin\mu$.
$\GF$ implicitly includes the residual effects of radiative corrections
not explicitly included in equation \ref{equ:radcorlept}.


The partial widths for the decays $\tau^-\rathin\mathrm{h}^-\nutau(\gamma)$, with
$\mathrm{h}^-=\pi^-/\mathrm{K}^-$, are given by
\begin{equation}
  \Gammahad   \equiv    \frac {\Bh} {\tau_\tau}
                 =      \frac {\GF^2 m_\tau^3} {16\pi} f_{\mathrm{h}}^2 
                        \radcorhad |V_{\alpha\beta}|^2 \psfachad,
         \label{equ:gammahad}
\end{equation}
where $f_{\mathrm{h}}$ are the hadronic form factors, $f_\pi$ and $f_{\mathrm{K}}$,
and $V_{\alpha\beta}$ are the CKM matrix elements, $V_{\mathrm{ud}}$ and $V_{\mathrm{us}}$,
for $\pi^-$ and $\kn$ respectively.
From an analysis of
$\pi^-\rathin\mu^-\bar{\nu}_\mu$ and
$\kn\rathin\mu^-\bar{\nu}_\mu$ decays, 
one obtains
$f_\pi |V_{\mathrm{ud}}| = \FPIVUDval$ and
$f_{\mathrm{K}} |V_{\mathrm{us}}| = \FKVUSval$\cite[and references therein]{MARCIANO92A}. 
The radiative correction factor, $\radcorhad$, is given by\cite{MARCIANO88A}                 
\begin{equation}
   \radcorhad   =  1 + \frac{2\alpha}{\pi} \mathrm{ln} \left( \frac {m_Z} {m_\tau} \right)+\cdots \simeq 1.02.
  \label{equ:radcorhad}
\end{equation}
The ellipsis represents terms, estimated to be ${\cal{O}}(\pm 0.01)$\cite{MARCIANO92A},        
which are neither explicitly treated nor implicitly absorbed into $\GF$,
$f_\pi |V_{\mathrm{ud}}|$, or $f_{\mathrm{K}} |V_{\mathrm{us}}|$.
For massless neutrinos the phase-space function, $\psfachad^0$, is given by
\begin{equation}
   \psfachad^0 (x) = (1 - x)^2
     \simeq   \cases {0.9877, & ($\mathrm{h}=\pi$); \cr
                      0.8516, & ($\mathrm{h}=\mathrm{K}$); \cr}
  \label{equ:psfachad}
\end{equation}
where $m_{\mathrm{h}}$ is the hadron mass and $x=m_{\mathrm{h}}^2 / m_\tau^2$.
%


We now consider the effects of neutrino masses and mixing on the $\tau$ decay rates.
The expression for $\Gammalept$, allowing for non-zero neutrino masses 
and mixing between $n$ lepton generations, is given by:
\begin{equation}
  \Gammalept    =   \frac {\GF^2 m_\tau^5}{192\pi^3} \radcorlept
                    \sum_{i=1}^{n} 
                    \sum_{j=1}^{n} 
                   |U_{\tau{i}}|^2 |U_{\ell{j}}|^2  
                   \psfaclept,
\label{equ:gammaleptnew}
\end{equation}
%
%
%
where $U_{ai}$ is the lepton mixing matrix\cite[page 276]{PDG96} 
and $\psfaclept$, the phase-space factor, depends on the neutrino and 
charged lepton masses.

The electron and muon neutrinos couple dominantly to $\nu_1$ and $\nu_2$  
and have small masses\cite{PDG96}.
%
%
We therefore neglect the masses of $\nu_1$ and $\nu_2$ and their mixing with $\nu_3$.
We do, however, consider the possible existence of a fourth generation  
neutrino, $\nu_{\mathrm{L}}$, of mass $m_{\nu_{\mathrm{L}}}>M_{\mathrm{Z}}/2$ 
which mixes with $\nu_\tau$
(LEP data have constrained the number of neutrinos of mass  
$m_{\nu}<M_{\mathrm{Z}}/2$ to be $N_\nu=2.991\pm0.016$\cite{PDG96}).
Since such a neutrino is kinematically forbidden in $\tau$ decays, 
the corresponding phase-space factor is zero.
$\Gammalept$ is therefore suppressed by a factor which depends on 
the strength of the mixing of the third and fourth generations, 
but not on the mass $m_{\nu_{\mathrm{L}}}$.
From the above, equation~\ref{equ:gammaleptnew} reduces to
\begin{equation}
  \Gammalept =   \frac {\GF^2 m_\tau^5}{192\pi^3} 
                    \radcorlept
                    (\mixingfactor)
                    \psfaclept,
\label{equ:gammaleptfinal}
\end{equation}
where the factor of $(\mixingfactor)\equiv|U_{\tau3}|^2$ allows for the 
Cabibbo-like suppression due to fourth generation mixing, 
and the phase-space factor is given by:
\begin{equation}
   \psfaclept (x, y) = \psfaclept^0(x)- 8y(1-x)^3+\cdots
   \label{equ:psfaclept2} 
\end{equation}
where $y=m_{\nu_3}^2 / m_\tau^2$,
$\psfaclept^0$ is given by equation \ref{equ:psfaclept} and the ellipsis
represents negligible terms of higher order in $m_{\nu_3}$.
%
%
Figures~{\fignopsfac}(a) and {\fignopsfac}(b) show the variation of 
$\psfaclept/\psfaclept^0$ with $m_{\nu_3}$ for
$\tau^-\rathin\mathrm{e}^-\bar{\nu}_{\mathrm{e}}\nutau$ and 
$\tau^-\rathin\mu^-\bar{\nu}_\mu\nutau$ decays respectively.

The expression for $\Gammahad$, allowing for non-zero neutrino masses and mixing 
in a similar fashion, is given by:
\begin{eqnarray}
  \Gammahad  & = & \frac {\GF^2 m_\tau^3} {16\pi} f_{\mathrm{h}}^2 \radcorhad |V_{\alpha\beta}|^2  
                       \sum_{i=1}^{n} |U_{\tau{i}}|^2  
                       \psfachad \nonumber \\
             & = & \frac {\GF^2 m_\tau^3} {16\pi} f_{\mathrm{h}}^2 \radcorhad |V_{\alpha\beta}|^2
                       (\mixingfactor)
                       \psfachad
         \label{equ:gammahadnew}
\end{eqnarray}
where the phase-space function has been calculated to be: 
\begin{equation}
   \psfachad (x, y) =
   \psfachad^0 (x)  
   \left[ 1 - y \left( \frac{2+x-y} {1-x} \right) \right]
   \sqrt{ 1 - y \left( \frac{2+2x-y} {(1-x)^2} \right)}, 
   \label{equ:psfachad2}
\end{equation}
with $y = m_{\nu_3}^2 / m_\tau^2$.
Figures~{\fignopsfac}(c) and {\fignopsfac}(d) show the variation of 
$\psfachad/\psfachad^0$ with $m_{\nu_3}$ for
$\tau\rathin\pi^-\nutau$ and
$\tau\rathin\mathrm{K}^-\nutau$ decays respectively.
Note the lower sensitivity to $m_{\nu_3}$ of the (two-body) hadronic decay modes 
compared to the (three-body) leptonic modes, 
despite the higher masses of the final state hadrons.
\section{Constraints on ${\mathbf{\nu_\tau}}$ mass and mixing}
In the standard model of electroweak interactions
the three lepton generations interact in an identical way,
apart from effects caused by their differing masses. 
In particular, they each couple with the same strength to the charged weak current.
If this assumption of universality is relaxed for the $\tau$, 
then the $\GF^2$ factor appearing in equations 
\ref{equ:gammalept} and \ref{equ:gammahad} may be replaced by 
$\GF\GFtau$, where $\GFtau$ is not necessarily equal to $\GF$.
If the measured values of $\GFtau$, evaluated assuming massless neutrinos and 
no mixing, appeared to be significantly smaller than $\GF$
it could indicate either new physics\cite{MARCIANO95A} or the suppression 
of decay rates due to non-zero neutrino masses or mixing.

We use the Particle Data group values and errors\cite{PDG96} for the 
measured quantities, in particular:
\begin{center}
\begin{tabular}{rcl cc rcl}                                 
  $\tau_\tau$  & $=$ & $\TAUTval$; & & & $m_\tau$         & $=$ & $\MTval$;\\
  $\Be$        & $=$ & $\BRTEval$; & & & $\Bm$            & $=$ & $\BRTMval$;\\                     
  $\Bp$        & $=$ & $\BRTPval$; & & & $\Bk$            & $=$ & $\BRTKval$. 
%
%
\end{tabular}
\end{center}
The sole exception is the use of the BES $\tau$ mass\cite{MTAUBESNEW}, 
obtained from the measurement of $\tau^+\tau^-$ production near threshold, 
since this has no dependence on $m_{\nu_\tau}$.
Substituting in equations \ref{equ:gammalept} and \ref{equ:gammahad} for 
the measured quantities, and assuming that $m_{\nu_\tau}=0$ and $\sin^2\theta_{\mathrm{L}}=0$ 
we obtain
\begin{equation}
\frac {\GFtau} {\GF} = 
\cases {\GFRATEval, & for $\tau^-\rathin\mathrm{e}^-\bar{\nu}_{\mathrm{e}}\nu_\tau$; \cr
        \GFRATMval, & for $\tau^-\rathin\mu^-       \bar{\nu}_\mu\nu_\tau$;          \cr
        \GFRATPval, & for $\tau^-\rathin\pi^-\nu_\tau$;                              \cr
        \GFRATKval, & for $\tau^-\rathin\kn  \nu_\tau$.                              \cr}
\end{equation}
where the number of standard deviations from the universality prediction (of unity) are shown in parentheses.
%
These results, which are all consistent with unity, indicate 
that the lepton couplings are universal and 
show no indications for non-zero neutrino mass or mixing. 
We therefore {\em{assume}} universality holds and use the 
measured $\tau$ partial widths to constrain $m_{\nu_3}$ and $\sin^2\theta_{\mathrm{L}}$.

We have mapped the likelihood of observing the measured $\tau$ partial widths,
as a function of both $m_{\nu_3}$ and $\sin^2\theta_{\mathrm{L}}$, 
by randomly sampling all the quantities used according to their experimental errors, 
allowing for the estimated 1\% theoretical uncertainty on $\radcorhad$.
The CLEO measurement of the $\tau$ mass was used to further constrain $m_{\nu_3}$.
From an analysis of 
$\tau^+\tau^-$ 
$\rightarrow$ 
$(\pi^+n\pi^0\bar{\nu}_\tau)$
$(\pi^-m\pi^0\nu_\tau)$ 
events (with $n\leq2, m\leq2, 1\leq n+m\leq3$), CLEO determined the $\tau$ mass to be 
$m_\tau = (1777.8 \pm 0.7 \pm 1.7) + [m_{\nu_3}(MeV)]^2/1400 MeV$\cite{CLEOWEINSTEIN}.
The likelihood for the CLEO and BES measurements to agree, as a function of 
$m_{\nu_3}$ is included in the global likelihood. 

Figure \figprobs(a)\ shows the 90\% and 95\% contours of the two dimensional 
likelihood distribution combined for all four $\tau$ decay channels.
No evidence is seen for a non-zero neutrino mass, nor for mixing.
By integration of the two dimensional likelihood over all values of 
$\sin^2\theta_{\mathrm{L}}$ we obtain the one-dimensional likelihood for $m_{\nu_3}$, 
independent of the value of $\sin^2\theta_{\mathrm{L}}$, 
as shown by the solid line of figure \figprobs(b). 
We obtain the following upper limits: 
$m_{\nu_3}<\MNUTCvalintA(\MNUTCvalintB)$\,{\em{MeV}} at the 90(95)\% confidence levels.
The solid line of figure \figprobs(c) shows the one-dimensional likelihood distribution for 
$\sin^2\theta_{\mathrm{L}}$, integrated over all values of 
$m_{\nu_3}$, from which we derive the upper limits:
$\sin^2\theta_{\mathrm{L}}<\FMIXCvalintA(\FMIXCvalintB)$ at the 90(95)\% confidence levels.
Table \tabnolimits\ summarises the upper limits of $m_{\nu_3}$ and $\sin^2\theta_{\mathrm{L}}$,
individually for each channel and for all channels combined.
Since the mixing of $m_{\nu_3}$ with other neutrinos is small, the limits derived for 
$m_{\nu_3}$ can be reasonably interpreted as limits on $m_{\nu_\tau}$.

Non-inclusion of the constraint from the CLEO $\tau$ mass measurement results 
in the following limits:  
$m_{\nu_3}<\MNUTAvalintA(\MNUTAvalintB)$\,{\em{MeV}} and 
$\sin^2\theta_{\mathrm{L}}<\FMIXAvalintA(\FMIXAvalintB)$
at the 90(95)\% confidence levels.
The slight reduction in the $\sin^2\theta_{\mathrm{L}}$ limit values, despite the  
increase in the $m_{\nu_3}$ limit, is due to the 
anticorrelation of the effects on the likelihood of non-zero values for
$m_{\nu_3}$ and $\sin^2\theta_{\mathrm{L}}$. 
%
%
%
%
%

%
%
\section{Conclusions}

We have derived the following constraints on the tau neutrino mass and fourth generation 
mixing from an analysis of the partial widths of tau lepton decays:
\mbox{$m_{\nu_\tau}<\MNUTCvalintA$\,{\em{MeV}}}  and 
\mbox{$\sin^2\theta_{\mathrm{L}}<\FMIXCvalintA$} at the 90\% confidence level.
These results are statistically and systematically independent of
traditional end-point analyses using multi-hadronic decays of the $\tau$.
Moreover we simultaneously consider mixing, and are insensitive to
fortuitous or pathological events close to the kinematic limits,
details of the resonant structure of multi-hadron $\tau$ decays, and
the absolute energy scale of the detectors.

These results will improve with more precise measurements of the $\tau$ mass, 
lifetime, and branching fractions, for example at a $\tau$-charm factory.
Ultimately, we expect that this method will be limited by the 
uncertainty on the $\tau$ lifetime. 
The extension of this technique to include multi-hadronic $\tau$ decays,
in conjunction with an improved theoretical description, should provide considerably 
more sensitivity due to the higher multiplicity of the final states.

%
%
%
%
%
%
%
\clearpage
\bibliography{/user/taylorl/tex/bib/general,/user/taylorl/tex/bib/tau_expt,/user/taylorl/tex/bib/tau_theory}
\clearpage
\noindent{\Large\bf{Table captions}} \\
 
\noindent{\bf{Table  \tabnolimits.}} \\
Upper limits obtained on $m_{\nu_3}$ and $\sin^2\theta_{\mathrm{L}}$, seperately for each $\tau$ decay 
channel and for all channels combined.
\vspace*{20mm}\\

\noindent{\Large\bf{Figure captions}} \\
 
\noindent{\bf{Figure \fignopsfac.}} \\
Relative phase-space suppression factors, as a function of tau neutrino mass for
             (a) $\tau^-\rathin\mathrm{e}^-\bar{\nu}_{\mathrm{e}}\nutau$,
             (b) $\tau^-\rathin\mu^-\bar{\nu}_\mu\nutau$,
             (c) $\tau^-\rathin\pi^-\nutau$, and
             (d) $\tau^-\rathin\kn\nutau$.
\vskip 10mm
 
\noindent{\bf{Figure \figprobs.}} \\
Likelihood distributions for all $\tau$ decay channels combined, for 
(a) $m_{\nu_3}$ {\em{vs.}} $\sin^2\theta_{\mathrm{L}}$,
(b) $m_{\nu_3}$, integrated over $\sin^2\theta_{\mathrm{L}}$, and 
(c) $\sin^2\theta_{\mathrm{L}}$, integrated over $m_{\nu_3}$.

\vskip 10mm
\clearpage
 
\begin{center}
\begin{tabular}{|l|crcr|crcr|}                                             \cline{2-9}
\multicolumn{1}{l} { }                                                 &
\multicolumn{4}{|c|}{Upper limit on $m_{\nu_3}$ ({\em{MeV}})}          &
\multicolumn{4}{|c|}{Upper limit on $\sin^2\theta_{\mathrm{L}}$}       \\   
\multicolumn{1}{l}{ }                                   &
\multicolumn{4}{|c|}{at the 90\% (95\%) C.L.}           &
\multicolumn{4}{|c|}{at the 90\% (95\%) C.L.}           \\ \hline
$\tau^-\rathin\mathrm{e}^-\bar{\nu}_{\mathrm{e}}\nutau$ &~~~~& \MNUTEvalintA & & (\MNUTEvalintB)~~~~ &~~~~ & \FMIXEvalintA & & (\FMIXEvalintB)~~~~ \\
$\tau^-\rathin\mu^-\bar{\nu}_\mu\nutau$                 &~~~~& \MNUTMvalintA & & (\MNUTMvalintB)~~~~ &~~~~ & \FMIXMvalintA & & (\FMIXMvalintB)~~~~ \\
$\tau^-\rathin\pi^-\nutau$                              &~~~~& \MNUTPvalintA & & (\MNUTPvalintB)~~~~ &~~~~ & \FMIXPvalintA & & (\FMIXPvalintB)~~~~ \\
$\tau^-\rathin\kn\nutau$                                &~~~~& \MNUTKvalintA & & (\MNUTKvalintB)~~~~ &~~~~ & \FMIXKvalintA & & (\FMIXKvalintB)~~~~ \\ \hline
%
%
All channels                                            &~~~~& \MNUTCvalintA & & (\MNUTCvalintB)~~~~ &~~~~ & \FMIXCvalintA & & (\FMIXCvalintB)~~~~ \\ \hline
%
%
\end{tabular}
\end{center}
 
\vskip 10mm

\begin{center}
{\Large\bf{Table \tabnolimits.}}
\end{center}

\clearpage
\begin{figure} [!htbp]
\begin{center}
    \mbox{\epsfig{file=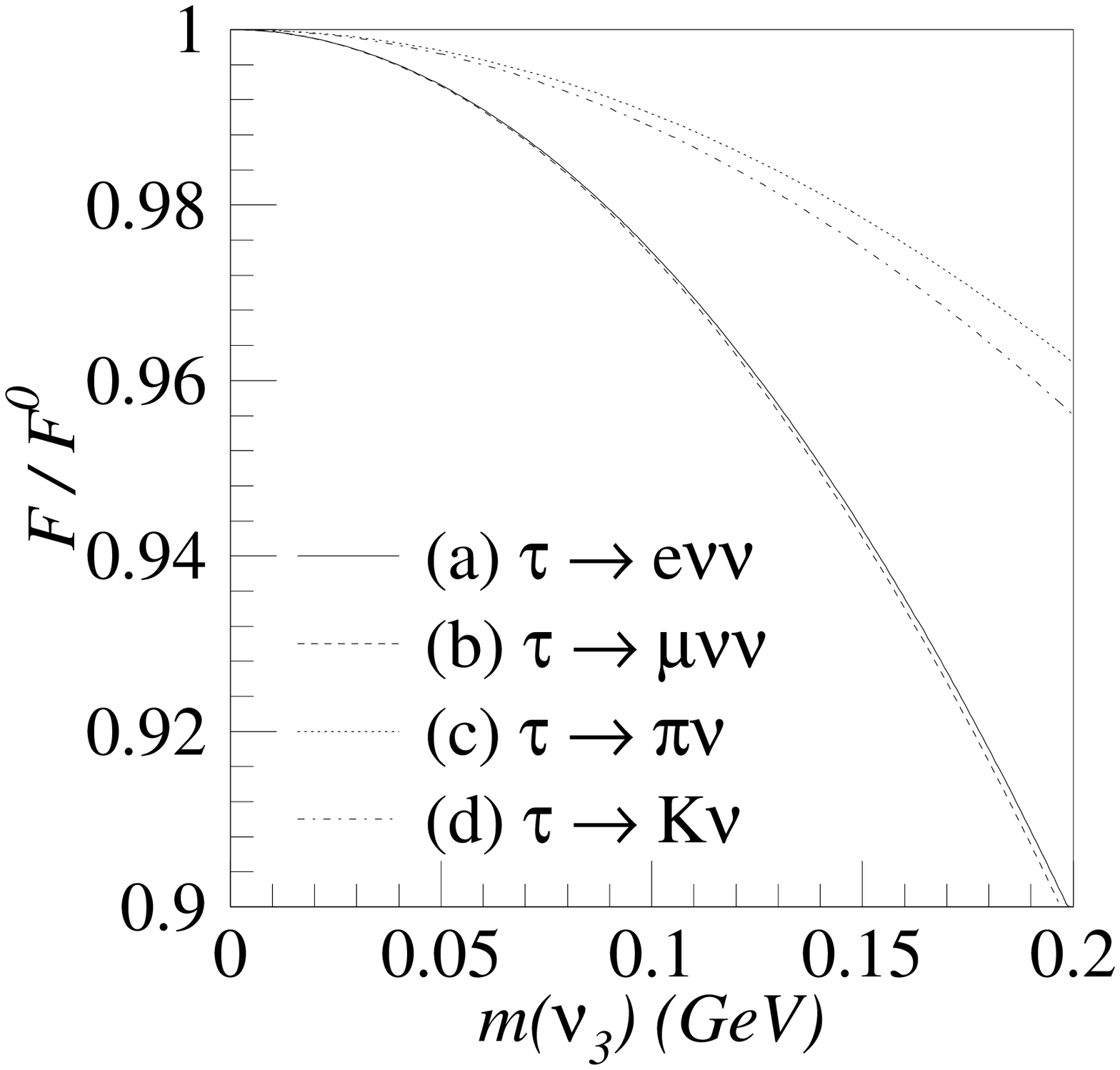,width=0.9\textwidth,clip=}\label{fig:psfacr}}
\end{center}
\end{figure}
 
\begin{center}
{\Large\bf{Figure \fignopsfac.}}
\end{center}
%
\clearpage
\begin{figure} [!htbp]
\begin{center}
    \mbox{\epsfig{file=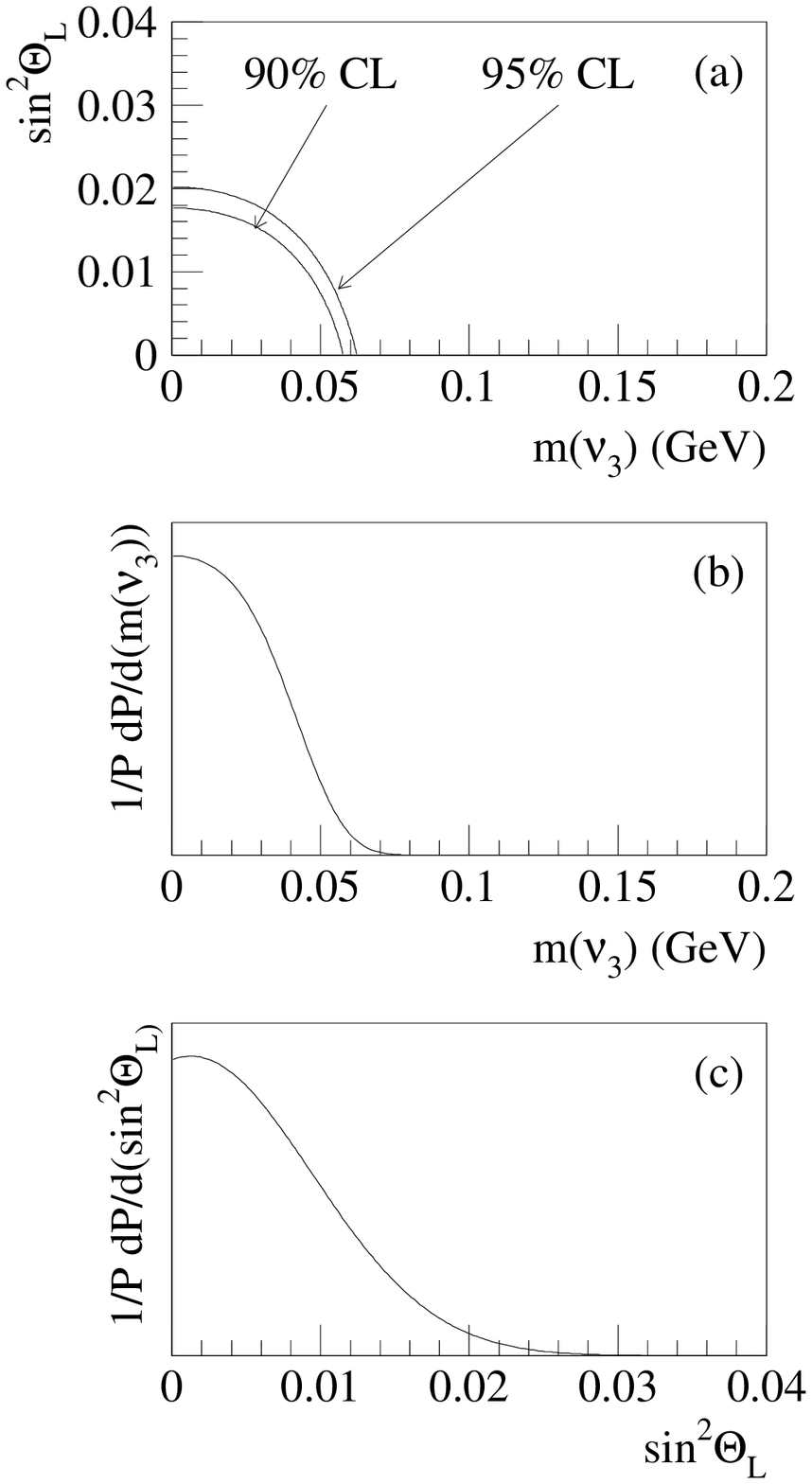,width=0.6\textwidth,clip=}\label{fig:contour}}
\end{center}
\end{figure}

\begin{center}
{\Large\bf{Figure \figprobs.}}
\end{center}
%
\end{document}

%% file: nutmass_include.tex
\renewcommand {\kn}{{\mathrm{K}^-}}

\newcommand {\Be}{{\cal{B}}_{\mathrm{e}}}
\newcommand {\Bm}{{\cal{B}}_{\mu}}
\newcommand {\Bp}{{\cal{B}}_{\pi}}
\newcommand {\Bk}{{\cal{B}}_{\mathrm{K}}}
\newcommand {\Bl}{{\cal{B}}_{\ell}}
\newcommand {\Bh}{{\cal{B}}_{\mathrm{h}}}
\newcommand {\Gammalept}{\Gamma_{\ell}}
\newcommand {\psfaclept}{F_{\ell}}
\newcommand {\radcorlept}{R_{\ell}}
\newcommand {\Gammahad}{\Gamma_{\mathrm{\mathrm{h}}}}
\newcommand {\psfachad}{F_{\mathrm{\mathrm{h}}}}
\newcommand {\radcorhad}{R_{\mathrm{\mathrm{h}}}}
\newcommand {\mixingfactor}{1-\sin^2\theta_{\mathrm{L}}}
\newcommand {\MTval}     {(1776.96^{+0.18+0.25}_{-0.21-0.17})\mathrm{MeV}}   
\newcommand {\TAUTval}   {(291.0      \pm 1.5)         \mathrm{fs}}    
\newcommand {\BRTEval}   {(17.80      \pm 0.08)        \%}             
\newcommand {\BRTMval}   {(17.30      \pm 0.10)        \%}             
\newcommand {\BRTPval}   {(11.07      \pm 0.18)        \%}             
\newcommand {\BRTKval}   {(0.71       \pm 0.05)        \%}             
\newcommand {\GF}        {G_{\mathrm{F}}}

\newcommand {\GFtau}     {G_{\mathrm{F}}^\tau}    
\newcommand {\GFMUval}   {(1.16639  \pm 0.00002) \times 10^{-5}   \mathrm{GeV^{-2}}} 
\newcommand {\GFRATEval} {0.996\pm0.007~~(-0.54\sigma)}      
\newcommand {\GFRATMval} {0.996\pm0.008~~(-0.55\sigma)}
\newcommand {\GFRATPval} {1.010\pm0.017~~(+0.60\sigma)}
\newcommand {\GFRATKval} {0.987\pm0.070~~(-0.18\sigma)}
\newcommand {\FPIVUDval} {(127.4 \pm 0.1)  \mathrm{MeV}}     %
\newcommand {\FKVUSval}  {(35.18 \pm 0.05) \mathrm{MeV}}     %
\newcommand {\MNUTEvalintA}     {63}
\newcommand {\MNUTMvalintA}     {66}
\newcommand {\MNUTPvalintA}    {141}
\newcommand {\MNUTKvalintA}    {333}
\newcommand {\MNUTAvalintA}     {59} 
\newcommand {\MNUTCvalintA}     {42} 
\newcommand {\MNUTEvalintB}   {72}
\newcommand {\MNUTMvalintB}   {76}
\newcommand {\MNUTPvalintB}  {160}
\newcommand {\MNUTKvalintB}  {375}
\newcommand {\MNUTAvalintB}   {68} 
\newcommand {\MNUTCvalintB}   {48} 
\newcommand {\FMIXEvalintA}    {0.016}
\newcommand {\FMIXMvalintA}    {0.018}
\newcommand {\FMIXPvalintA}    {0.028}
\newcommand {\FMIXKvalintA}    {0.036}
\newcommand {\FMIXAvalintA}    {0.013} 
\newcommand {\FMIXCvalintA}    {0.014} 
\newcommand {\FMIXEvalintB}    {0.019}
\newcommand {\FMIXMvalintB}    {0.022}
\newcommand {\FMIXPvalintB}    {0.033}
\newcommand {\FMIXKvalintB}    {0.038}
\newcommand {\FMIXAvalintB}    {0.016} 
\newcommand {\FMIXCvalintB}    {0.017} 


%% file: nutmass.bbl
\begin{thebibliography}{10}

\bibitem{MASSIVENU}
F.~Boehm and P.~Vogel.
\newblock {\em Physics of Massive Neutrinos}.
\newblock Cambridge University Press, Cambridge, 1987.

\bibitem{DARKMATTER}
D.~Seckel J.~R.~Primack and B.~Sadoulet.
\newblock {\em Ann. Rev. Nucl. Part. Sci.}, 38:751, 1988.

\bibitem{SOLARDEFICITA}
S.~P. Rosen and J.~M. Gelb.
\newblock {\em Phys. Rev.}, D34:969, 1986.

\bibitem{SOLARDEFICITB}
H.~A. Bethe.
\newblock {\em Phys. Rev. Lett.}, 56:1305, 1986.

\bibitem{ALEPHNUTAU}
D.~Buskulic et~al.
\newblock {\em Phys. Lett.}, B349:585, 1995.

\bibitem{SHROCK81B}
R.~E. Shrock.
\newblock {\em Phys. Rev.}, D24:1275, 1981.

\bibitem{BRYMAN87A}
D.~A. Bryman and C.~E. Picciotto.
\newblock {\em Phys. Rev.}, D36:3514, 1987.

\bibitem{SAMUEL88A}
M.~A. Samuel and R.~R. Mendel.
\newblock {\em Mod. Phys. Lett}, A3(No. 4):393, 1988.

\bibitem{SHROCK81A}
R.~E. Shrock.
\newblock {\em Phys. Rev.}, D24:1232, 1981.

\bibitem{SHARMA84A}
R.~R.~L. Sharma and N.~K. Sharma.
\newblock {\em Phys. Rev.}, D29:1533, 1984.

\bibitem{RAJPOOT88A}
S.~Rajpoot and M.~A. Samuel.
\newblock {\em Mod. Phys. Lett}, A3(No. 16):1625, 1988.

\bibitem{LI91A}
Xue-Qian Li and Tao Zhi-jian.
\newblock {\em Phys. Rev.}, D43:3691, 1991.

\bibitem{MARCIANO95A}
W.~J. Marciano.
\newblock {\em Nucl. Phys. B (Proc. Suppl.)}, 40:3, 1995.

\bibitem{BERMAN58A}
S.~M. Berman.
\newblock {\em Phys. Rev.}, 112:267, 1958.

\bibitem{KINOSHITA59A}
T.~Kinoshita and A.~Sirlin.
\newblock {\em Phys. Rev.}, 113:1652, 1959.

\bibitem{SIRLIN78A}
A.~Sirlin.
\newblock {\em Rev. Mod. Phys.}, 50:573, 1978.

\bibitem{MARCIANO88A}
W.~J. Marciano and A.~Sirlin.
\newblock {\em Phys. Rev. Lett.}, 61:1815, 1988.

\bibitem{PDG96}
{Particle Data Group}.
\newblock {\em Phys. Rev.}, D54:1, 1996.

\bibitem{MARCIANO92A}
W.~J. Marciano.
\newblock {\em Phys. Rev.}, D45:R\,721, 1992.

\bibitem{MTAUBESNEW}
J.Z. Bai.
\newblock {\em Phys. Rev.}, D53:20, 1996.

\bibitem{CLEOWEINSTEIN}
R.~Balest et~al.
\newblock {\em Phys. Rev.}, D47:R3671, 1993.
\newblock The neutrino mass dependence quoted in this paper is in error due to
  a typographical oversight (A. Weinstein, private communication). This does
  not, however, affect any other numbers in the paper.

\end{thebibliography}
